\documentclass[aps,prl,reprint,superscriptaddress,apsrev]{revtex4-1}
\usepackage{empheq}
\usepackage{graphicx}
\usepackage{epsfig}
\usepackage{amsmath,amsthm}
\usepackage{color}
\usepackage{verbatim}
\usepackage{ulem}
\begin{document}
%
\title{Disorder-induced gap in the normal density of states of the organic superconductor $\kappa$-(BEDT-TTF)$_2$Cu[N(CN)$_2$]Br}
\author{Sandra Diehl}\email{sandradiehl@uni-mainz.de}
\affiliation{Graduate School Materials Science In Mainz, Staudingerweg 9, 55128 Mainz, Germany}
\affiliation{Institut f\"ur Physik, Johannes Gutenberg-Universit\"at Mainz, Staudingerweg 7, 55128 Mainz, Germany}
\author{Torsten Methfessel}
\affiliation{Institut f\"ur Physik, Johannes Gutenberg-Universit\"at Mainz, Staudingerweg 7, 55128 Mainz, Germany}
\author{Jens M\"uller}
\author{Michael Lang}
\author{Michael Huth}
\affiliation{Physikalisches Institut, Goethe-Universit\"at, Max-von-Laue-Str.\ 1, 60438 Frankfurt (M), Germany}
\author{Martin Jourdan}
\author{Hans-Joachim Elmers}
\affiliation{Institut f\"ur Physik, Johannes Gutenberg-Universit\"at Mainz, Staudingerweg 7, 55128 Mainz, Germany}
\date{\today}
\begin{abstract}
The density of states of the organic superconductor $\kappa$-(BEDT-TTF)$_2$Cu[N(CN)$_2$]Br, measured by scanning tunneling spectroscopy on \textit{in-situ} cleaved surfaces, reveals a logarithmic suppression near the Fermi edge persisting above the critical temperature $T_\mathrm{c}$. 
A soft Hubbard gap as predicted by the Anderson-Hubbard model for systems with disorder exactly describes the experimentally observed 
suppression.
The electronic disorder also explains the diminished coherence peaks of the quasiparticle density of states below $T_\mathrm{c}$.\\

\end{abstract}
%
\maketitle

The anomalous states of matter which may evolve near correlation-driven metal-insulator transitions continue to be of high interest in condensed matter physics. Prime examples are the unusual metallic and unconventional superconducting states which have been found in the proximity of Mott insulating states in systems such as transition metal oxides or organic charge-transfer salts  \cite{McKenzie1997,Imada1998,Georges2004}. Here, electron localization, and the associated charge gap, is the consequence of a strong on-site Coulomb repulsion. An aspect of considerable recent interest in this field of research relates to the interplay of correlation effects and disorder. One may ask whether an approach that puts disorder more into focus may help for better understanding some of the intriguing electronic properties of these complex materials where electron correlations and disorder inevitably coexist.

While there has been considerable effort from the theoretical side on this issue~\cite{Altshuler1980, Fukuyama1980,Dobrosavljevic2003,Shinaoka2009,Shinaoka2009a,Byczuk2009}, only a few experimental investigations have been reported so far~\cite{Kim2005,Sano2010,Sasaki2012}. In particular, to our knowledge, there is no study on the most fundamental aspect of how disorder affects the single particle density of states for a material close to the Mott transition.
  
In this Letter, we apply scanning tunneling spectroscopy (STS) to a model substance, the organic charge-transfer salt $\kappa$-(BEDT-TTF)$_2$Cu[N(CN)$_2$]Br (in short $\kappa$-Br), to study the electronic density of states (DOS) for a correlated metal and superconductor in the presence of disorder. This material has been known to reside on the metallic side of a bandwidth-controlled Mott metal insulator transition~\cite{Kanoda1997,Toyota2007}, where cooling through a glass-like structural transition at $T_g = 75 - 80$\,K gives rise to a certain amount of disorder~\cite{Su1998,Muller2002,Yoneyama2004,Toyota2007,Hartmann2014}.
In fact, the material’s low-temperature properties, such as dc resistivity~\cite{Limelette2003} and optical conductivity~\cite{Merino2008}, were found to be consistent with the predictions by dynamical mean field theory (DMFT) for a correlated metal~\cite{Merino2000}, implying an enhanced density of states (DOS) at the Fermi energy $E_F$. However, no indications for a corresponding quasi-particle peak have been reported so far. 

Through simultaneous topographic studies on in-situ cleaved surfaces, we prove that the measurements have been performed on a well-defined system. In the present paper the main focus lies on the energy dependence of the DOS in the normal state, complemented by signatures of the superconducting state which serve as a characterization of the crystal under investigation. The salient result here is the observation of a logarithmic correction to the normal-state DOS predicted by various theories. Our data are in good agreement with the predictions of the Anderson-Hubbard model \cite{Shinaoka2009,Shinaoka2009a} which aims at describing electronically correlated systems with short-range Coulomb interactions in the presence of a spatially uncorrelated, spin-independent random potential.

%
%

Single crystals of $\kappa$-Br were grown in an electrochemical crystallization process \cite{Wang1990}. Since as-grown crystals are not suited for surface-sensitive STS measurements, we developed an \textit{\textit{in-situ}} cleaving technique for obtaining clean crystal surfaces.
The crystal was cleaved parallel to the $ac$-plane and then attached to an Omicron sample holder using conducting adhesive. After \textit{in-situ} cleaving at a pressure of \mbox{$1\times 10^{-10}$}\,mbar the sample was mounted into the pre-cooled scanning tunneling microscope (STM) operating at a base pressure of about \mbox{$5 \cdot 10^{-11}$}\,mbar. 
By this procedure the cooling rate changed from initially $- 3.5$\,K/min to a value of $-1$\,K/min 
at the glass-transition temperature $T_g \sim 75 - 80$\,K. This implies a residual disorder in the orientational degrees of freedom of the ET molecules' terminal ethylene groups of order $\sim 3$\,\% \cite{Hartmann2014}. For the spectroscopic investigations we used an electrochemically etched tungsten tip flashed at about 2200\,K. We measured $I\left(V\right)$ curves in the temperature range from 5\,K to 13\,K, averaged over at least 100 curves and differentiated them numerically to receive differential conductivity $dI/dV$ spectra. This method avoids erroneous energy shifts one may otherwise obtain when using fast scans. A comparison of data taken with different set points before the feed back loop was switched off confirmed that the $dI/dV$ spectra did not depend on the absolute tunneling current $I$, thus excluding influences from local sample heating or a possible local suppression of superconductivity caused by high current densities.
%
%
%
\begin{figure}[t]
\includegraphics[width=1.0\columnwidth]{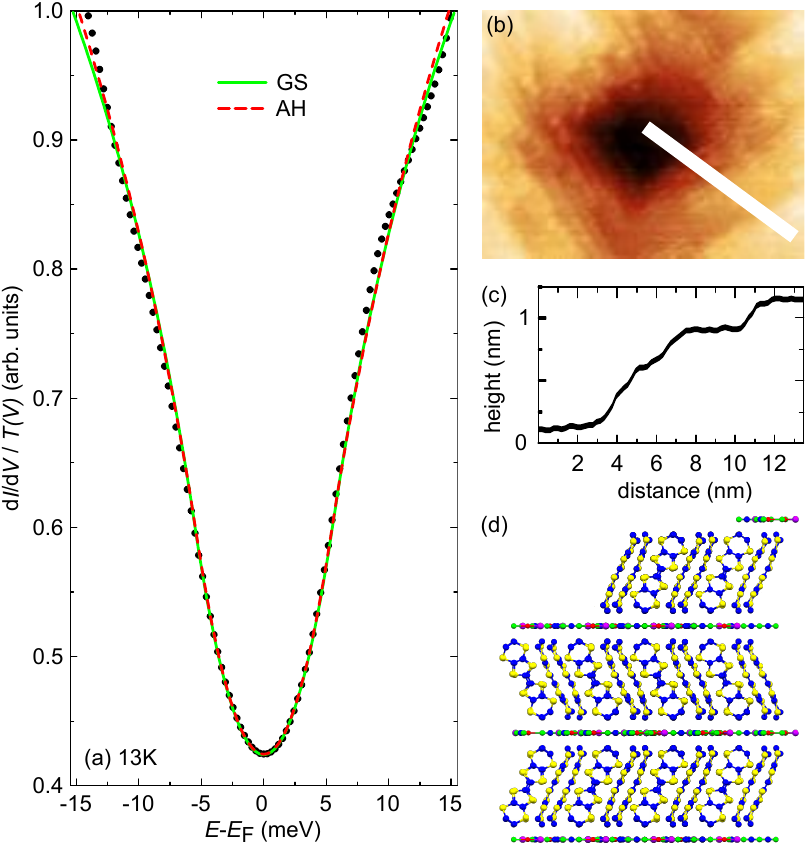}
\caption{
(a) d$I$/d$V$ spectrum normalized to $T(V)$ of the $\kappa$-(BEDT-TTF)$_2$Cu[N(CN)$_2$]Br crystal measured at 13\,K. The initial tunneling current is $I$=100\,pA and the initial bias voltage is $V$=30\,mV. The data is shown as black dots. The dark gray area indicates $|V|<V_0$.
The red full line describes the fit to $B(V)$. 
(b) 23$\times$18nm$^2$ STM image of the cleaved surface. 
(c) Height profile along the line in (b). 
(d) Corresponding sketch of the crystal surface indicating two different step heights.
}
\label{fig1}
\end{figure}

In Fig.~\ref{fig1}(b) a representative STM image of the cleaved surface is shown revealing step heights (Fig.~\ref{fig1}(c)) 
of the anion (0.3~nm) and cation (0.9~nm) layers (see schematic in Fig.~\ref{fig1}(d)), 
where the sum is slightly smaller than the bulk layer distance of 1.5~nm, indicating surface relaxation. d$I$/d$V$ spectra taken at anion and cation surface positions proved to be similar.

In STS, the spectra have to be corrected for the different work functions of tip and sample which leads to a voltage-dependent asymmetric tunneling transmission function $T(V)$ \cite{Ukraintsev1996}. For this reason all spectra shown here are normalized to $T(V)$ which was obtained from spectra taken at 13\,K, {\it i.e.}, in the normal state of $\kappa$-Br. We furthermore assume a constant DOS for the tip material in the relevant energy range so that the $dI/dV/T(V)$ curves reflect the thermally-smeared DOS $D(V)$ of the sample with $eV=E-E_\mathrm{F}$ ($E_F$ denotes the Fermi energy).
In Fig.~\ref{fig1}(a) we present a spectrum at 13\,K in the normal state showing a significant reduction at zero bias. In the following we analyze this data in the framework of the Anderson-Hubbard (AH) model discussed by Shinaoka and Imada \cite{Shinaoka2009,Shinaoka2009a} for disordered itinerant electron systems with short-range interactions. 

\begin{figure}[t]
\includegraphics[width=0.9\columnwidth]{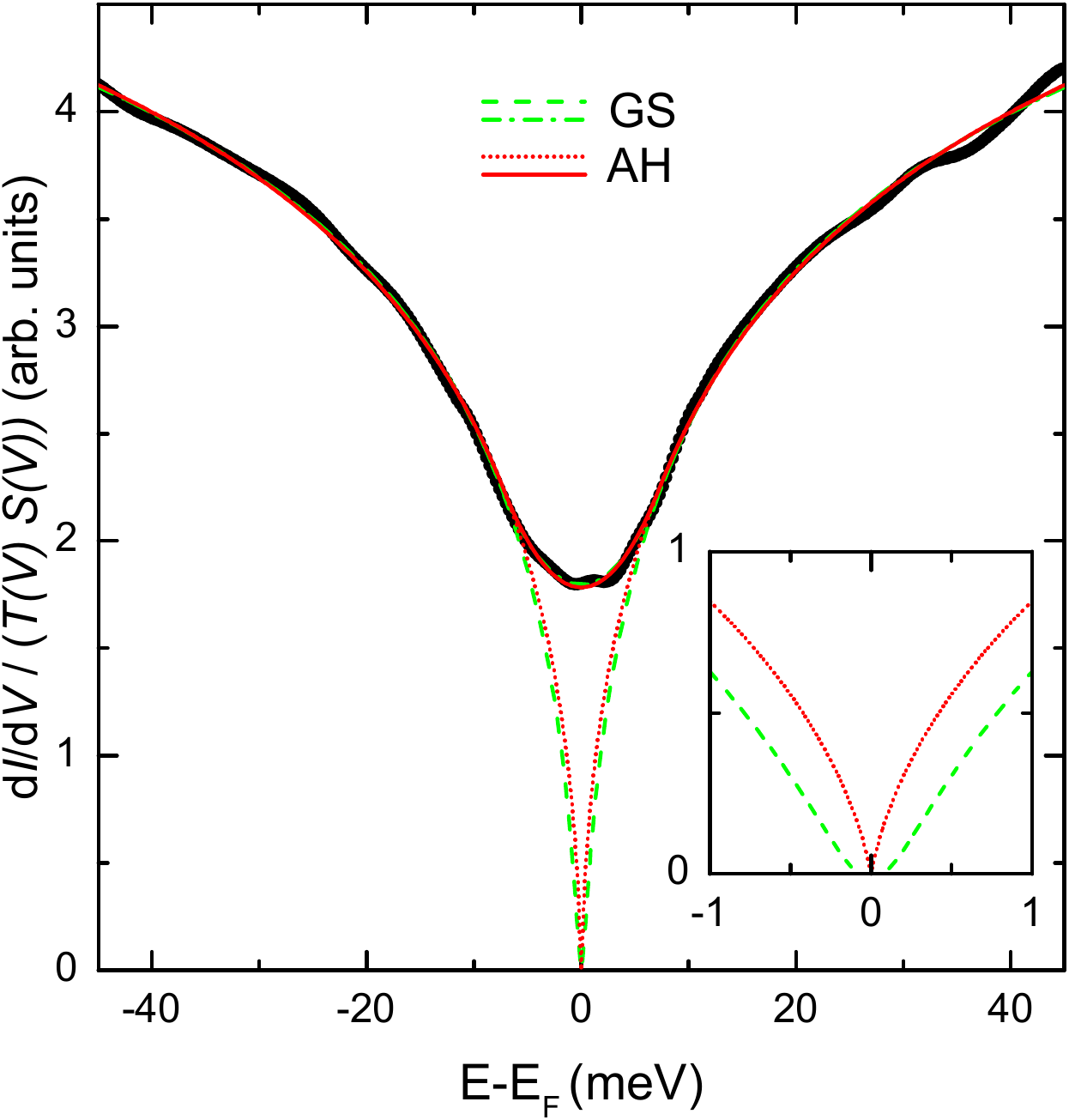}
\caption{\label{fig:B(V)}Comparison of the AH~\cite{Shinaoka2009} and 
the granular-electronic-system model~\cite{Beloborodov2004} with experimental data (dots). The dashed and the dotted lines show the behavior for zero temperature whereas the solid and dash dotted lines  respect a thermal activation according to Eq.~\eqref{eq:cosh}.
The inset shows a magnified region close to $E=E_{\rm F}$.}
\end{figure}

In their numerical treatment of the AH model a scaling law is introduced for the DOS in the presence of short-range Coulomb interactions and a multi-valley energy landscape \cite{Shinaoka2009}:
\begin{equation}
B_{AH}(V)=\exp \left[ -\alpha \left(- \log \left|\mathrm{e}V\right|\right)^d \right],  \, \left|V\right|\geq V_0\, ,
\label{eq:Imada}
\end{equation}
where $d$ denotes the spatial dimension and $\alpha$ is a non-universal constant.  
We obtain an excellent fit to our data in Fig.~\ref{fig1}(a) setting $d=2$ and $\alpha=0.288$. We note, however, that the fit fails to describe the data in a small energy range of about $5 - 8$\,meV around zero-bias, denoted as $\mathrm{e}V_0$ in Tab.~\ref{tab:delta}. This becomes apparent in Fig.~\ref{fig:B(V)} where we show an extrapolation of the AH-fitted DOS for small bias voltages. Phenomenologically, an excellent fit of this region is obtained assuming a hard energy gap of small size and employing a DOS function of the following form (index {\it HG} for {\it hard gap}):
\begin{equation}
B_{HG}(V)= c  \cosh\left(\frac{\mathrm{e}V}{\varepsilon_T}\right) , \, \left|V\right| < V_0,
\label{eq:cosh}
\end{equation}
where $c$ is a constant
and $\varepsilon_T$ measures the effective barrier height. For finite temperatures $T>0$\,K, $B_{HG}(T)$ rapidly becomes non-zero near $E_\mathrm{F}$  because a thermally-activated crossing of the small barrier $\varepsilon_{T}$ leads to a nearly temperature-independent 
prefactor and the voltage-dependence is described by the $\cosh$-term. With the requirement of continuous differentiability at the inflection point $V_0$, the $c(T)$ and $\varepsilon_T$ terms are fixed for any given temperature and do not represent adjustable parameters.

For comparison, one may consider an alternative description, also giving rise to a logarithmic correction to the DOS. This model, proposed by Beloborodov et al.~\cite{Beloborodov2003,Beloborodov2004}, describes granular electronic systems. Here the charge transport occurs between spatially localized regions of enhanced density of itinerant states which are coupled via tunneling processes. With increasing strength of the tunnel coupling $g$, an insulator-to-metal transition occurs. For strong coupling ($g \geq 1$) this leads to an electronic state that resembles in several respects a disorder-modified Fermi liquid and for which the term granular Fermi-liquid was coined \cite{Beloborodov2004}.  For a two-dimensional system in the limit of large coupling strength one finds (index GS for granular system):
\begin{equation}
B_{GS}(V)=\gamma \Bigg(1-\frac{1}{4 \pi g_T^0}  \ln\left(\frac{ g_T^0 E_{C}}{\left|\mathrm{e}
	V\right|}\right)\Bigg)^{4 g_T^0}, \, \left|V\right|\geq V_0,
\label{eq:Coulomb}
\end{equation}
with $\gamma =c_0 \left({g_T^0 E_C}/{\left|\mathrm{e} V\right|} \right)^{{1}/{\pi}}$. $g_T^0$ denotes the bare tunnel conductance in units of 2e$^2/\hbar$. Our fits result in a temperature-independent value of $g_T^0=1.0\pm0.1$. $E_{C}$ denotes the charging energy of a region of high electron density. The constant $\gamma$ is proportional to the DOS for noninteracting electrons. In this model, the suppression of the DOS near $E_\mathrm{F}$ is a consequence of the Coulomb interaction, as the capacitance $C$ leads to a charging energy $E_C=\mathrm{e}^2/2 C$ which hinders charge transport for $|eV|<E_C$ (for $T$\,=\,0\,K).
The fit to our data at 13\,K, shown as green dashed line in Fig.~\ref{fig1}(a), describes the data well for $V>V_0$ whereas for $V<V_0$ we again use the phenomenological hard gap DOS as discussed above (see also Fig.~\ref{fig:B(V)}). However, we note that the deduced capacitance of the regions of high electron density is of the order of $C=(1.0\pm 0.3) \cdot 10^{-20}$\,F. For an assumed spherical region this value corresponds to a characteristic radius \mbox{$r \approx 1$}\,{\AA}. Such a small length scale calls the applicability of this model into question, as it implies the discretization of energy levels in these regions, which, however, is not explicitly taken into account in the model assumptions \cite{Beloborodov2004}.\\ 
We therefore argue that the AH model provides a more reasonable description of the logarithmic energy dependence of the DOS observed here. 
At the same time, we would like to point out that this type of disorder-induced gap is predicted by both theories and
may represent a general feature of correlated-electron systems on the metallic side of a metal-insulator transition in the presence of weak disorder.

\begin{figure}[t]
\includegraphics[width=0.9\columnwidth]{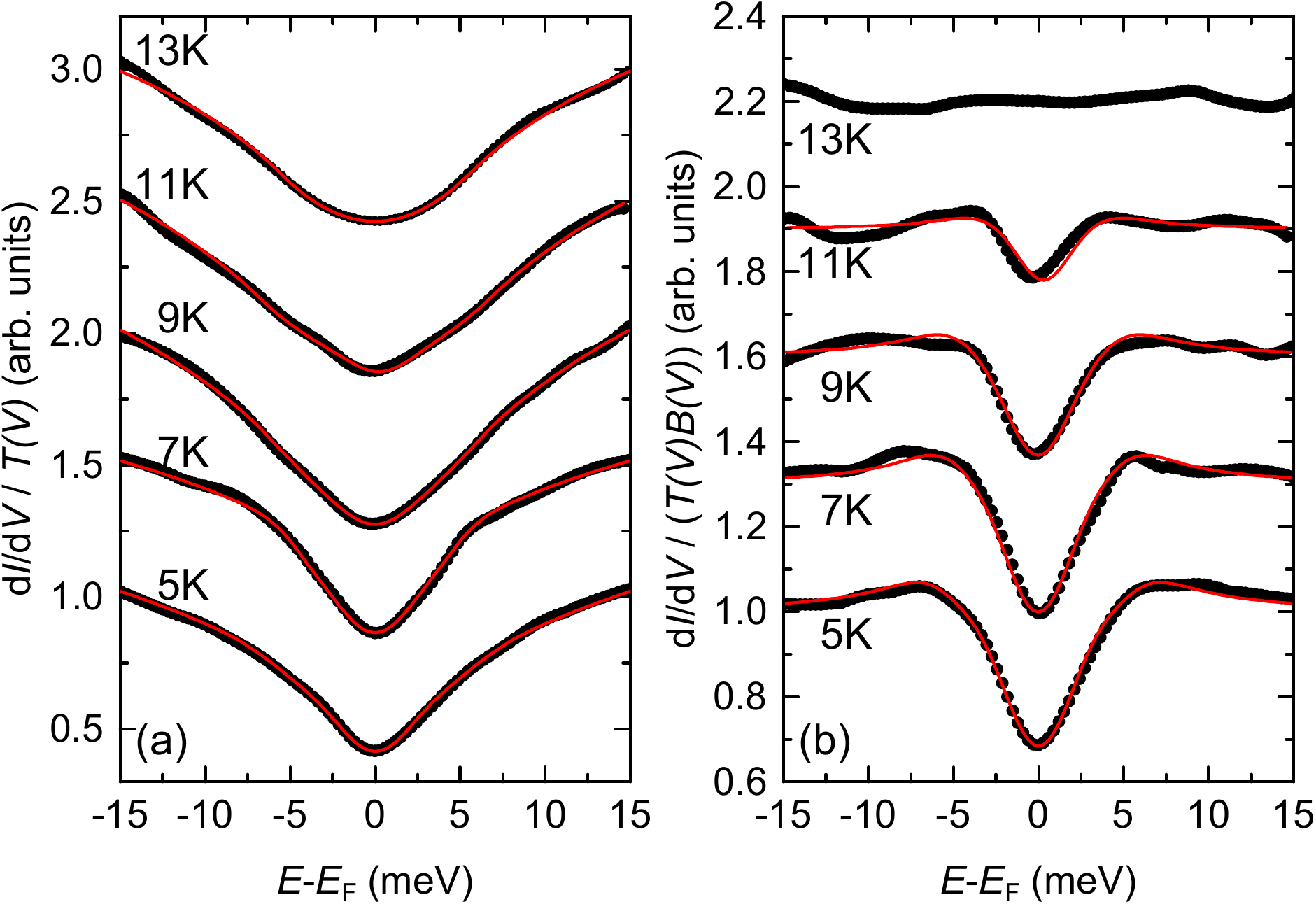}
\caption{ Temperature dependence of the $\mathrm{d}I/\mathrm{d}V$ spectra. The dots show the measured data and the solid lines represent the corresponding fit. (a) Spectra normalized to $T(V)$ and fitted with $B(V) S(V)$. (b) Spectra normalized to $T(V) B(V)$  and fitted with $S(V)$.}
\label{fig3}
\end{figure}

We now turn to the DOS signatures of the superconducting state of $\kappa$-Br. In order to obtain a complete description of the energy dependence of the DOS for the whole investigated temperature range from 5\,K to 13\,K, we use a Dynes fit for the single-particle DOS $S(V)$ for $T<T_c$ \cite{Dynes1978}:
\begin{multline}
S\left(V\right) = \int_{-\infty}^{\infty}\frac{\partial f\left(E+\mathrm{e}V,T\right)}{\partial V} \cdot 
\\
 \left| \operatorname{Re} \left[ \frac{\mathrm{e}V + \mathrm{i}\mathit{\Gamma}}{\sqrt{\left(\mathrm{e}V + \mathrm{i}\mathit{\Gamma}\right)^2 - \mathit{\Delta}^2 }}\right]\right|\, dE \quad .
\label{eq:Dynes}
\end{multline}
Accordingly, the DOS is described by $D(V)=T(V)B(V)$ with $B(V)=B_{AH}(V)$ or $B(V)=B_{GS}(V)$ for $T>T_c$ and $V>V_0$, and by $D(V)=B(V)S(V)$ for $T<T_c$. In Fig.~\ref{fig3} we show the fits obtained for all spectra in the range $5 - 13$\,K employing the fit parameters collected in Tab.~\ref{tab:delta}. For completeness, we used for these fits both model descriptions for the normal DOS as no significant difference in the quality of the fits using either $B(V)=B_{AH}(V)$ or $B(V)=B_{GS}(V)$ was noticeable.

$S(V)$ accurately describes the experimental data [see Fig.~\ref{fig3}\,(b)].
We stress that the comparison of fits with $d$-wave and $s$-wave symmetry does not reveal any significant differences.
In particular, characteristic features, such as a V-shaped DOS expected for order parameters with $d$-wave symmetry, are absent. However, the observed spectra are only an indirect test for the symmetry of the order parameter as the analysis requires an integration of 
of the Fermi surface in $k$-space~\cite{Weiss1999,Goddard2004,Toyota2007}, see supplemental information \cite{SI}.

Figure~\ref{fig:deltavstemp} shows the temperature dependence of the gap size $\mathit{\Delta}$ and the 
broadening parameter $\mathit{\Gamma}$ resulting from the fit to the experimental data, see {Ref.~\onlinecite{SI}} for the fit parameters at different temperatures. 
The ratio $2\mathit{\Delta}(T=0\,\mathrm{K}) /k_\mathrm{B} T_\mathrm{c} = 5{.}28\pm 0{.}13$ is in excellent agreement with the value of 5.4 obtained from specific heat measurements \cite{Elsinger2000}, whereas earlier STS results revealed significantly enhanced values of 6.7 \cite{Arai2001_2} and 9 \cite{Bando1990}. 
The large value obtained here, as compared to the weak-coupling BCS value of $2\mathit{\Delta}(T=0\,\mathrm{K})=3{.}52 k_\mathrm{B} T_\mathrm{c}$, along with the close proximity of superconductivity and an antiferromagnetic insulating state suggests a strong-coupling magnetic mechanism.
We stress that the resulting $\mathit{\Gamma}$ is large compared to values observed for other superconductors, as {\it e.g.}
MgB$_2$ ($\Gamma \sim 0.2$\,mV)~\cite{Iavarone2002} comprising grains of size $50 - 500$\,nm.
We attribute this observation to the disorder-induced broadening of the DOS in the superconducting state. The Dynes function thus represents an effective description of the DOS whereas its spatial inhomogeneity gives rise to an enhanced $\mathit{\Gamma}$ value.
With the increase of fluctuations upon approaching the superconducting transition, we expect that the electronic disorder becomes 
less effective resulting in the observed decrease of $\mathit{\Gamma}$.

\begin{table}[t]
\caption{fit parameters resulting from a fits to Eqs.~\eqref{eq:Imada}--\eqref{eq:Coulomb} to the experimental data.}
\label{tab:delta}
\begin{tabular}{ccccc}
\hline
	T 		& $E_C$		& e$V_0$ 	& $ \varepsilon_T$	& $\alpha$	\\ 
	(K)		& (eV)		            & (meV)	& (meV)		&		\\ \hline
	5		& 12(5)		& 8(1)		& 0.10(3)	&	0.220(1)		\\
 	7		& 12(5)		& 5(1)		& 0.16(3)	&	0.201(2)		\\
 	9		& 19(5)		& 6(1)		& 0.18(3)	&	0.339(3)		\\
 	11		& 21(5)		& 7(1)		& 0.15(3)	&	0.328(4)		\\
	13		& 13(1)		& 6(1)		& 0.17(3)	&	0.288(3)		\\ \hline
\end{tabular}
\end{table}

\begin{figure}[t]
\includegraphics[width=0.8\columnwidth]{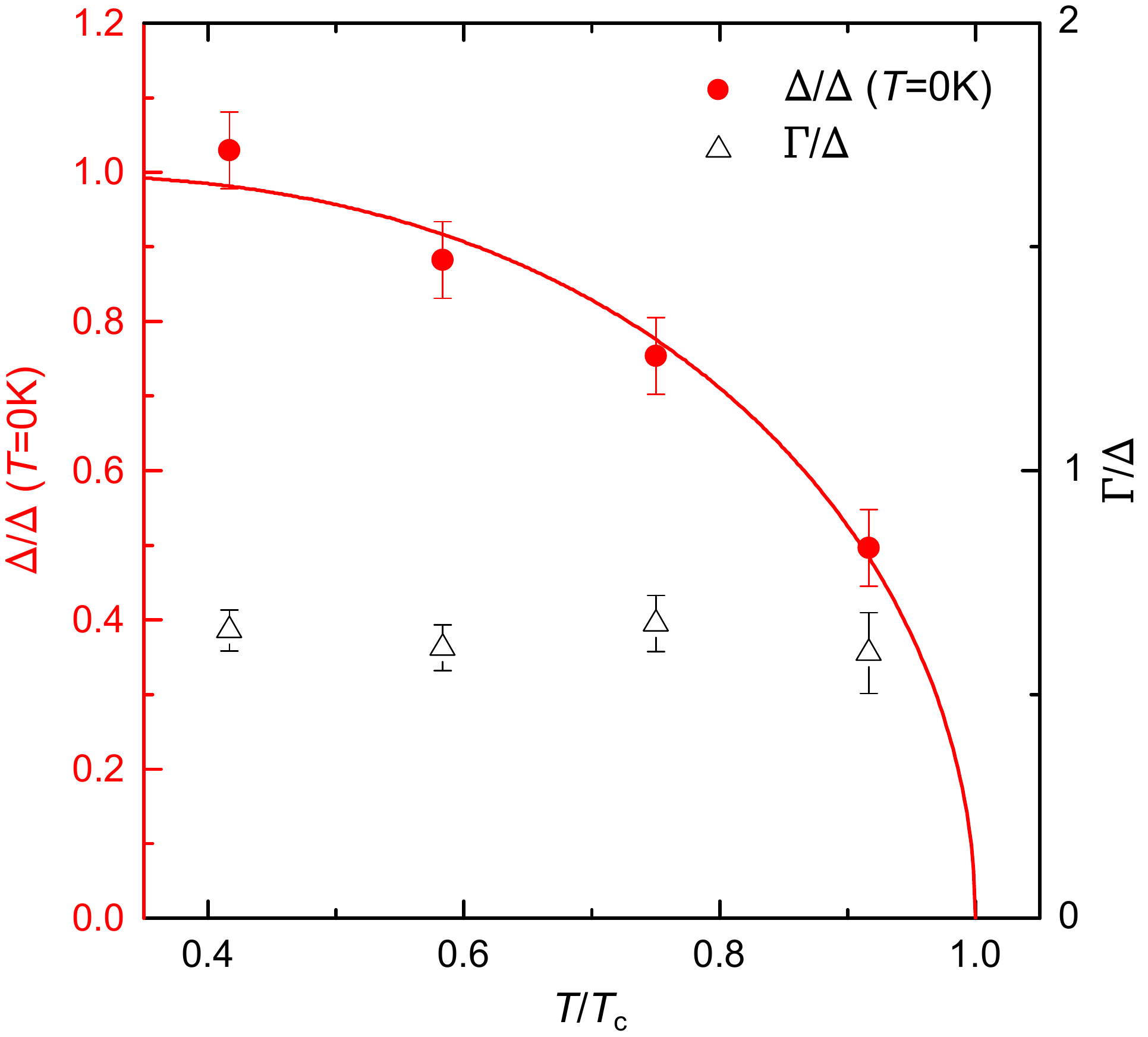}
\caption{\label{fig:deltavstemp} Left Scale: Gap width ($d$-wave) $\mathit{\Delta}_1$ normalized to the value at $T=0$\,K in dependence of the temperature normalized to $T_\mathrm{c}$ with $\mathit{\Delta}(T=0\,\mathrm{K}) = 2.72$\,meV and $T_\mathrm{c} = 12.0$\,K. Right Scale: Ratio of the broadening $\mathit{\Gamma}$ to the gap width $\mathit{\Delta}$. 
}
\end{figure}

%
In summary, measurements of the temperature-dependent differential-conductivity spectra of the organic superconductor $\kappa$-Br reveal a logarithmic suppression of the DOS at the Fermi energy, which persists above the superconducting transition temperature $T_\mathrm{c}$.
By a careful analysis of the spectra, in particular by a comparison of different logarithmic correction terms to the DOS predicted by theory, we are able to fit our data to the energy-dependent DOS functions based on the AH and granular-electronic-system model. 
The latter results in an unreasonably short length scale involved, seemingly incompatible with the model assumptions. Such inconsistencies are absent for the AH model.
Our experiment thus strongly supports the validity of this model for $\kappa$-Br.\\
On the basis of fits of the normal DOS we are able to unfold the DOS in the superconducting state. Here we find, relying on a simple life-time broadening analysis of the single particle spectrum \cite{Dynes1978}, an extrapolated zero-temperature energy gap $\mathit{\Delta}(T=0\,\mathrm{K})$ of 2.7\,meV and a BCS ratio $2\mathit{\Delta}(T=0\,\mathrm{K})/k_{\mathrm{B}}T_\mathrm{c}$ of 5.3,
which identifies $\kappa$-Br as a strong-coupling superconductor. Remarkably, the broadening parameter $\mathit{\Gamma}$ amounts to about 60\,\% of the zero-temperature gap size, consistent with previous observations~\cite{Ichimura2003_2}. 
We consider this to be a consequence of the disorder-induced renormalization of the normal DOS of $\kappa$-Br.\\
The logarithmic correction observed here for $\kappa$-Br may be characteristic for a broad class of correlated electron systems close to
a metal-insulator transition in the presence of disorder.

%
%
We thank the Deutsche Forschungsgemeinschaft (SFB/TR 49) and the Graduate School Materials Science in Mainz for financial support.



%

\end{document}